\documentclass[twocolumn,preprintnumbers,amsmath,amssymb,pra,floatfix]{revtex4}
\newcommand{\be}{\begin{equation}}
\newcommand{\ee}{\end{equation}}
\usepackage{graphicx}
\usepackage{multirow}
\graphicspath{{./data/},{./figures/}}

\begin{document}

\title{
Spin noise at an arbitrary spin temperature
}
\author{S.-K.~Lee}
\email{lsk@princeton.edu}
\affiliation{Physics Department, Princeton University, Princeton, NJ 08544, USA}
\date{\today}

\begin{abstract}
An ensemble of spins oriented along the $z$ direction exhibits nonzero fluctuation in the transverse ($x$- and $y$-) components of the spin angular momentum in accordance with the uncertainty principle. When the spins obey a spin temperature distribution, the mean square fluctuation in $S_{x}$ can be calculated by ensemble average of the expectation value of $S_{x}^{2}$ with respect to an equilibrium density matrix $\rho =e^{\beta S_{z}}/Z$. The fluctuation can also be calculated from the fluctuation-dissipation theorem as has been done in literature in the context of NMR spin noise. For spin 1/2 particles in the high temperature limit, appropriate for many NMR experiments, the two methods are known to produce the same, temperature-independent spin noise. We show that inclusion of the zero-point fluctuation term in the original Nyquist relation extends this correspondence to an arbitrary spin temperature for any spin $S$. This indicates that the uncertainty principle-limited spin projection noise can be viewed as a result of the zero-point fluctuation in the thermal bath coupled to the spins.
\end{abstract}

\maketitle
\smallskip

Spin noise is routinely measured in modern NMR spectrometers \cite{Hoult2001} and has important  consequences in certain precision measurements based on weak response of the spins to an applied field \cite{Bialek1986, Budker2006}. For spins at equilibrium with relatively small thermal polarization, the noise spectrum is often derived by applying the
fluctuation-dissipation theorem to the complex susceptibility of the spin ensemble driven by a transverse magnetic field. The resulting, temperature-independent noise for paramagnetic spin-1/2 ensembles \cite{Hoult2001, Budker2006} at first sight seems to fully reflect the fundamental quantum nature of the spin noise. However, in view of the high temperature approximation used in these calculations, the role of quantum fluctuations, enforcing quantum uncertainty in spin projections at any temperature, is not clear.

Spin projection noise is important in experiments involving optically pumped spins with high polarization and low spin temperature. The sensitivity of alkali atomic magnetometers, for example, is fundamentally limited by such noise. In this work we show that the zero-point fluctuation term in the Nyquist noise, usually neglected in thermal noise calculations, correctly produces the ``quantum part'' of the noise required by the uncertainty principle.

For a spin $\vec{S}$ fully polarized along $\hat{z}$, the uncertainty principle dictates that the rms uncertainty in the transverse components of the spin satisfies
\[
\Delta S_{x}\Delta S_{y}\ge \frac{1}{2}\left| \left<S_{z}\right>\right| =S/2.
\]
Without spin squeezing, which we do not consider in this work, the minimum
variance in $S_x$ is therefore
\be
(\Delta
S_{x})^{2}=S/2.
\label{eq1}
\ee

\smallskip

For an ensemble described by a density matrix
\[
\rho =\exp (\beta S_{z})/Z,
\]
where $\beta $ is the dimensionless inverse spin temperature and $Z$ is the partition function, the variance is
\begin{eqnarray*}
(\Delta S_{x})^{2} &=&\overline{\left\langle S_{x}^{2}\right\rangle
-\left\langle S_{x}\right\rangle ^{2}}=\text{Tr}(\rho S_{x}^{2}) \\
&=&\frac{1}{2}\left\{ \text{Tr}(\rho S_{x}^{2})+\text{Tr}(\rho S_{y}^{2})\right\}  \\
&=&\frac{1}{2}\text{Tr}(\rho (S^{2}-S_{z}^{2})) \\
&=&\frac{1}{2}S(S+1)-\frac{1}{2}\cdot \frac{\sum_{m}m^{2}e^{\beta m}}{%
\sum_{m}e^{\beta m}}\text{ }.
\end{eqnarray*}
The summation in the last term runs over the possible projection quantum
number $m=S,S-1,\cdots ,-S$.

The evaluation of the above equation is simplified if one observes that
\[
\frac{\sum_{m}m^{2}e^{\beta m}}{\sum_{m}e^{\beta m}}=\frac{1}{Z}\frac{%
\partial ^{2}Z}{\partial \beta ^{2}}=\left( \frac{1}{Z}\frac{\partial Z}{%
\partial \beta }\right) ^{2}+\frac{\partial }{\partial \beta }\left( \frac{1%
}{Z}\frac{\partial Z}{\partial \beta }\right)
\]
and
\[
\frac{1}{Z}\frac{\partial Z}{\partial \beta }=\frac{\partial }{\partial
\beta }\ln Z=SB_{S}(\beta )
\]
where
\[
B_{S}(x)\equiv \frac{1}{S}\,\,\left\{ \left( S+\frac{1}{2}\right) \coth
\left( S+\frac{1}{2}\right) x-\frac{1}{2}\coth \frac{x}{2}\right\}
\]
is the Brillouin function for angular momentum $S$ \cite{Reif1965}. It turns out that a Brillouin function satisfies the identity
\[
\frac{d}{dx}B_S(x)+S B_S^2(x) + \coth( x/2 ) B_S(x) = S+1
\]
from which
\be
\left( \Delta S_{x}\right) ^{2}=\frac{S}{2}\coth \frac{\beta }{2}\cdot
\,B_{S}(\beta).
\label{eq2}
\ee
Figure 1 shows the plot of $\left( \Delta S_{x}\right) ^{2}$ as a function of $\beta$ for a few values of $S$.

In the low temperature limit, we verify that
\[
\lim_{\beta \rightarrow \infty }\left( \Delta S_{x}\right) ^{2}\rightarrow
S/2
\]
and in the high temperature limit,
\[
\lim_{\beta \rightarrow 0}\left( \Delta S_{x}\right) ^{2}\rightarrow S(S+1)/3
\]
which states that the variances in all three components of the spin are equal
in this case. Also note that for $S=1/2$, $B_{S}(\beta )=\tanh \beta /2$ and
Eq. (2)\ reduces to $\left( \Delta S_{x}\right) ^{2}=1/4.$

\begin{figure}[b]
\includegraphics[width=9cm]{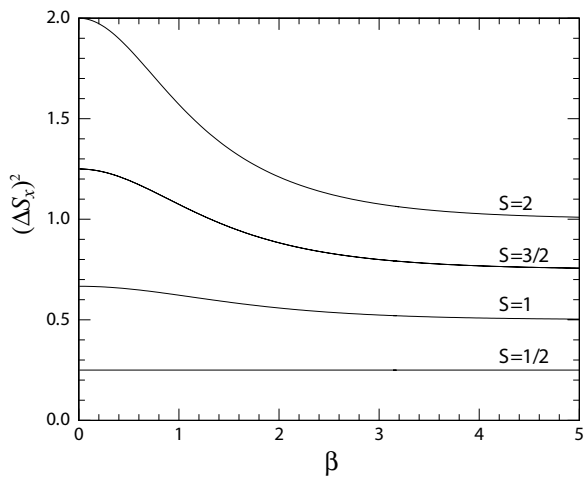}
\caption{Calculated mean square transverse spin fluctuation in a spin ensemble as a function of the spin temperature parameter $\beta$ for several values of $S$. }

\label{fig1}
\end{figure}

\smallskip

The transition from purely quantum mechanical spin projection noise to predominantly thermal spin noise is exactly reproduced by calculation based on the fluctuation-dissipation theorem. We  start by reviewing the line of argument presented in Ref.~\cite{Hoult2001}.
The method is the same as the one used for calculation of magnetic field noise from Johnson noise current in electrical conductors \cite{Lee2008}.

Suppose the spin ensemble is in spin-temperature equilibrium with a magnetic
moment $M_{0}\,$along the $z$ axis and is subject to a static magnetic field $%
B_{0}\hat{z}$. We emphasize that the derivation below does not require that $%
M_{0}$ be the thermal polarization in $B_{0}$; the equilibrium  distribution may be established in the presence of, for example, optical pumping and strong
spin-exchange interaction. Such equilibrium is parameterized by a
dimensionless coefficient in the Boltzmann factor, $\beta $, as is done
above. A temperature $T$ with unit of kelvin is then \textit{defined} using the Zeeman
splitting, $\hbar\omega_0\equiv\gamma\hbar B_0$, as $\beta \equiv \hbar \omega _{0}/kT$. Here $\gamma$ is the gyromagnetic ratio of the spins and $k$ is the Boltzmann constant. That such $T$ is physical is evidenced by adiabatic demagnetization cooling.

When a transverse continuous rf field is applied to the spins the energy dissipation
in the sample represents an effective resistive load to the driving coil, $%
R_{\text{NMR}}$. This resistance is related to the noise voltage in the coil when
the coil is {\it not} driving the spins,
\be
S_{V}(\omega )=4kTR_{\text{NMR}}
\label{eq3}
\ee
according to the conventional high-temperature form of the Nyquist noise. Here $%
S_{V}(\omega )$ is the noise voltage power spectrum. For spins with
transverse relaxation time $T_{2}$, $R_{\text{NMR}}$ is given by \cite{Hoult2001}
\[
R_{\text{NMR}}=\frac{1}{2}~\frac{\gamma \omega b_{1}^{2}T_{2}}{1+\left( \Delta
\omega \,\,T_{2}\right) ^{2}}M_{0}
\]
where $b_{1}$ is the field produced by a unit current in the coil and $%
\Delta \omega =\omega -\omega _{0\text{ }}\,$is the offset of the driving
frequency $\omega $ from $\omega _{0}$. We restrict
ourselves to the common NMR case where the linewidth $1/T_{2}$ is much
smaller than $\omega _{0}$. Then $\omega \,$in the numerator of $R_{\text{NMR}}$
can be replaced by $\omega _{0}$ in integrating $S_{V}(\omega )$ over $%
\omega $, which gives the mean square voltage noise
\[
\left( \Delta V\right) ^{2}=\int_{0}^{\infty }S_{V}(\omega )\frac{d\omega }{%
2\pi }=\gamma \omega _{0}b_{1}^{2}kTM_{0}\,\,.
\]
In the high-temperature limit, if the ensemble has $N$ spin 1/2 particles
with magnetic moment $\mu =\gamma \hbar /2$,
\be
M_{0}=N\mu \hbar \omega_{0}/2kT.
\label{eq4}
\ee

This leads to the well known temperature-independent spin noise voltage
\[
\left( \Delta V\right) ^{2}=\mu ^{2}\omega _{0}^{2}b_{1}^{2}N
\]
from which the mean square transverse magnetic moment per particle is $%
(\Delta \mu _{x})^{2}=\mu \,,$ as expected from a system with only two
possible measurements ($\mu _{x}=\pm \mu )$.

A generalization of the above derivation to an arbitrary spin $S$ at an
arbitrary temperature can be made by replacing the high temperature Nyquist
noise Eq. (3) and equilibrium polarization Eq. (4) as following.

\begin{eqnarray}
S_{V}(\omega )&=& 2\hbar \omega \,\coth \left( \hbar \omega
/2kT\right) \,\,R_{\text{NMR}}\,,\\
M_{0} &=& N\,\mu \,B_{S}(\hbar \omega _{0}/kT).
\end{eqnarray}

The first equation is a result of replacing average thermal energy $kT$ of a
bosonic mode by its quantum mechanical counterpart including zero-point
fluctuation \cite{Nyquist1928, Callen1951, Koch1982}, $kT\rightarrow \hbar \omega \,(\,1/2+1/(e^{\hbar \omega
/kT}-1)\,)=(\hbar \omega /2)\,\coth (\hbar \omega /2kT)$. Equation (6) is in fact the original definition of a Brioullin function \cite{Reif1965}. If we now assume once again sharp resonance when integrating $S_{V}(\omega )$, we obtain for
the spin noise voltage
\[
\left( \Delta V\right) ^{2}=\frac{1}{2S}\mu ^{2}\omega
_{0}^{2}b_{1}^{2}N\,\coth (\hbar \omega _{0}/2kT)\,B_{S}(\hbar \omega
_{0}/kT)\text{ .}
\]
Noting that $\hbar \omega _{0}/kT=\beta$, we get for the mean square transverse
magnetic moment per particle
\[
\left( \Delta \mu _{x}\right) ^{2}=(\hbar \gamma )^{2}\frac{S}{2}\,\coth
\frac{\beta }{2}\cdot \,B_{S}(\beta )
\]
which is exactly Eq. (2) scaled for a magnetic moment. This is our main result. For $S=1/2$, and only for this spin, the high-temperature approximation gives a result valid for all temperatures.

Whereas the density matrix calculation gives the total spin noise in the absence of dynamic information of the spin ensemble, calculation based on the fluctuation-dissipation theorem gives the spectral distribution of such noise. In the above derivation assumption of sharp resonance led to the approximation where the frequency dependence is completely determined by the Zeeman resonance. Extension to the case of low quality-factor Zeeman resonance is straightforward, where non-Lorentzian noise spectrum is obtained.

In an electrical circuit, the Nyquist noise spectrum including the zero-point fluctuation has been measured in resistively shunted Josephson junctions by Koch et al. \cite{Koch1982}. Corresponding test of spin noise as calculated in this work should readily be performed by employing, for example, hyperpolarized noble gas atoms with high nuclear spin polarization. Laser polarization of $^{129}$Xe, $^{131}$Xe, and $^{83}$Kr with $I$=1/2, 3/2, 9/2, respectively, is experimentally established.

It is interesting to consider the effect of the spin noise considered above on an atomic magnetometer based on the Zeeman resonance of alkali atoms. For such a magnetometer, an important contribution to the fundamental noise comes from the spin projection noise, as was calculated in Ref. \cite{Savukov2005}. There, the noise was calculated for fully polarized $^{39}$K atomic spins in their stretched state with total angular momentum along the quantization axis $\left<F_{z}\right>=2$. Our calculation indicates that the transverse spin noise is in general a function of the polarization with increased noise at low polarizations. For high-sensitivity rf magneotmeters demonstrated in Ref. \cite{Savukov2005, Lee2006}, however, high level of spin polarization corresponding to $\beta\gtrsim3$ is required for suppression of spin-exchange broadening. Therefore, given the weak dependence of spin noise on $\beta$ in this regime, additional spin noise in these magnetometers should be insignificant.

\vspace{12pt}
\begin{acknowledgments}
The author acknowledges helpful discussions with Prof. Michael Romalis.
\end{acknowledgments}

\end{document}